\documentclass[11pt,titlepage]{article}
\usepackage{setspace} 
\usepackage{amsmath}
\usepackage{mathtools}
\usepackage{amssymb}
\usepackage{graphicx}
\usepackage[round,numbers,sort&compress]{natbib}
\newcommand{\beq}{\begin{equation}}
\newcommand{\beqn}{\begin{eqnarray}}
\newcommand{\eeq}{\end{equation}}
\newcommand{\eeqn}{\end{eqnarray}}
\newcommand{\ra}{\rightarrow}
\newcommand{\lra}{\leftrightarrow}

\title{Requirements for efficient cell-type proportioning: regulatory timescales, stochasticity and lateral inhibition}

\author{Benjamin Pfeuty\thanks{Corresponding author.} \\
  Université de Lille, CNMRS, Laboratoire de Physique des Lasers, Atomes, et Mol\'ecules \\ 
  F-59000, Lille, France \\
  \and Kunihiko Kaneko \\
  Department of Pure and Applied  Sciences \\
  University of Tokyo, Tokyo 153-8902, Japan }

\begin{document}

\maketitle

\abstract{The proper functioning of multicellular organisms requires the robust establishment of precise proportions between distinct cell-types.
  This developmental differentiation process typically involves intracellular regulatory and stochastic mechanisms to generate cell-fate diversity as well as intercellular mechanisms to coordinate cell-fate decisions at tissue level.
  We thus surmise that key insights about the developmental regulation of cell-type proportion can be captured  by the modeling study of clustering dynamics in population of inhibitory-coupled noisy bistable systems.
This general class of dynamical system is shown to exhibit a very stable two-cluster state, but also frustrated relaxation, collective oscillations or steady-state hopping which prevents from timely and reliably reaching a robust and well-proportioned clustered state. 
To circumvent these obstacles or to avoid fine-tuning, we highlight a general strategy based on dual-time positive feedback loops, such as mediated through transcriptional versus epigenetic mechanisms, which improves proportion regulation by coordinating early and flexible lineage priming with late and firm commitment.
This result sheds new light on the respective and cooperative roles of multiple regulatory feedback, stochasticity and lateral inhibition in developmental dynamics.

}

\clearpage

\section*{Introduction}%

The development of multicellular organisms relies on sophisticated collective behaviors of interacting cells such as aggregation~\cite{Gregor2010}, segmentation~\cite{Soroldoni2014} or cell-type diversification~\cite{Chang2008}.
One intriguing collective phenomena has been termed canalization~\cite{Waddington1942} and refers to the developmental ability of a multicellular organism to produce the same end-result, such as the proportion between distinct cell types, regardless of variability of its environment or genotype.
Cell-type proportion is primarily the result of a sequential differentiation process wherein multipotent cells select between two distinct fate-restricted cell subtypes.
These binary cell-fate decisions are regulated by the interplay between intracellular regulatory mechanisms required to generate diverse cell-type attractors from a single precursor cell type and intercellular signaling mechanisms required to coordinate fate decisions in population of cells (Table~\ref{Table1}).
On the one hand, differentiation regulatory networks typically involve positive feedback loops such as self-activation, mutual activation or inhibition, or more elaborate motifs~\cite{Guantes2008}, which may also operate at different timescales~\cite{Brandman2005}.
On the other hand, cell-fate choice is influenced by intercellular coupling via juxtacrine or paracrine signaling, notably through a lateral inhibition mechanism whereby cells prevent each other from differentiating into the same type~\cite{Greenwald1992}.

The two systems where regulation of cell-type proportions have received the most attention are the multicellular development of the social amoeba Dictyostelium and early lineage specification during mammalian embryogenesis.
Vegetative Dictyostelium cells exposed to starving conditions aggregate to form slugs and, eventually, fruiting bodies, which contains spores and stalk cells with a relatively precise ratio of about 4:1~\cite{Bonner1949,Sakai1973}.
During this process, starving cells start to differentiate into a heterogeneous population of prespore or prestalk cells where prespore cells elicit Dif1 signals that promote other cells to acquire the stalk fate~\cite{Kay2001,Williams2006}.
The spore and stalk phenotypes are later stabilized at the mound and fruiting body stages, presumably though positive feedback generated by intracellular or autocrine signaling mechanisms~\cite{Katoh2004,Anjard2005}. 
Quite similar developmental features are observed in early mammalian embryos where the small pool of cells (16-32 cell stage) of the inner cell mass is equally segregated in two populations of primitive endoderm (Gata6-positive) and epiblast (Nanog-positive) lineage cells~\cite{Gardner1979}. 
Cells are first biased to a specific lineage in a reversible manner as Nanog-positive cells contribute to increase extracellular Fgf4 levels that influence Fgf4-bound cells by preventing accumulation of Nanog and promoting accumulation of Gata6~\cite{Frankenberg2011}. 
Cell fate is later stabilized at E4.5 in an irreversible manner~\cite{Grabarek2012}, presumably through the activation of stabilizing intracellular feedback mechanisms~\cite{Frankenberg2011,Bessonnard2014} and cell motility and sorting processes~\cite{Xenopoulos2015}.

In both developmental systems, cell-type diversification and proportioning relies on the interplay between intracellular regulatory mechanisms and intercellular inhibitory signals, and tends to occur as a two-step differentiation process. In contrast, positional information or cell division are not critical factors in regulating these developmental differentiation processes, though it can be the case for other developmental lineage decisions~\cite{Ralston2008,Kicheva2014}.
An important issue is then to identify which regulatory features are minimally required for efficient proportioning process~\cite{Kaneko1994,Mizuguchi1995,Vakulenko2009,Nakajima2008}.
To address this issue, we derive a model consisting in a population of inhibitory-coupled, noisy and bistable cells, which recapitulates the main intracellular and intercellular properties described above.
The rigorous and exhaustive analysis of this class of model allows to identify the major obstacles to efficient cell-type proportioning depending on the strengths of noise, intracellular feedback and intercellular coupling.
We further identify a universal strategy based on dual-time regulatory feedback loops to circumvent these obstacles. 
Finally, we discuss how this strategy is implemented in biological systems, notably through the dichotomy between transcriptional and epigenetic regulation, and how other mechanisms may further improve the cell-type proportioning process or coordinate it with growth or patterning processes.

\begin{table}[!ht]
  \centering     
  \begin{tabular}{|c|c|c|c|}
    \hline
    {\bf Cell lineages (organism)}     &  {\bf Positive feedback}      & {\bf Signaling}    & {\bf Ref} \\\hline
    Spore/Stalk (Dictyostelium)        &  Sdf1 $\lra$  DhkA            & Dif1$^{p}$  & \cite{Kay2001,Anjard2005} \\\hline
    Tricho-/Atricho-blast (Arabidopsis)&  Wer $\lra$ Gl3               & CPC$^{p}$         & \cite{Lee2002,Schellmann2002} \\\hline
    Anchor/Ventral uterine (C. Elegans)&  Lin12 $\lra$ Lin12           & Lin12-Lag2$^{j}$ & \cite{Wilkinson1994} \\\hline 
    Neuro-/epidermo-blast (Drosophila) &  AS-C $\vdash \dashv$ E(Spl)-C& Delta-Notch$^{j}$ & \cite{Heitzler1996} \\\hline
    Motile/Primary cilia  (Vertebrate) &  Foxj1 $\lra$ Rfx1            & Jagged-Notch$^{j}$  & \cite{Choksi2014}  \\\hline  
    Hypo-/Epi-blast (Mammal)           &  Gata6 $\vdash \dashv$ Nanog  & Fgf2-Fgfr$^{p}$    & \cite{Frankenberg2011} \\\hline
    Neuron/Neural stem (Mammal)        &  Ngns $\vdash \dashv$ Cyclins  & Delta-Notch$^{j}$ & \cite{Pfeuty2015} \\\hline
  \end{tabular}
  \caption{{\bf Intracellular and intercellular pathways for binary decision.}
    Examples of developmental fate decisions based on intracellular positive feedback loops and intercellular inhibitory coupling. $p/j$: paracrine/juxtacrine signaling.}
  \label{Table1}
\end{table} 

\section*{Methods}%

\subsection*{Effective one-dimensional model for binary cell fate decision}%

Bistable behaviors occurring in a wide range of physical systems are often studied as effective one-dimensional models. 
Although protein networks are generally characterized with a sophisticated signaling and regulatory architecture, the mechanism underlying cellular bistability typically relies on a core positive feedback loop, such as mutual inhibition or mutual activation between two proteins.
The dynamics of circuits featured with simple positive feedback architecture can potentially be reduced to a one-dimensional model by using timescale separation arguments.  
Adiabatic elimination of fast modes is straigthforward when a fast variable can be explicitely identified (e.g., mRNA), but can also be done by appropriate changes of variables.
This is for instance the case for the toggle-switch circuit where two protein species of concentration $p_A$ and $p_B$, are activated by some signal $s_{A,B}$, interacts through mutual inhibition of strength $\mu$, and are eventually subjected to self-activation of strength $\alpha$ (Fig. 1A):
\begin{subequations}
  \begin{align}
  \label{toggle1}
  \frac{dp_A}{dt} &= \frac{s_A+\alpha\, p_A^2}{1+\alpha\, p_A^2}-(1+\mu\,p_B^2)\,p_A  \\
  \label{toggle2}
  \frac{dp_B}{dt} &= \frac{s_B+\alpha\, p_B^2}{1+\alpha\, p_B^2}-(1+\mu\,p_A^2)\,p_B 
  \end{align}
\end{subequations}
This dynamical system, often used as a generic model for binary differentiation decisions~\cite{Huang2007,Wang2011}, shows a transition from monostability to bistability as $s_{A,B}$ increases.
In the fully symmetric case $s_A=s_B$, the symmetric state $p_A=p_B$ is destabilized through a pitchfork bifurcation (supercritical or subcritical depending on $\alpha$ and $\mu$) giving rise to two coexisting stable states ${\cal A}$ ($p_A>p_B$) and ${\cal B}$ ($p_B>p_A$).
The coexistence of two stable fixed points separated by a saddle point $\vec{p}_0$ defines a 1D invariant manifold denoted ${\cal M}$ that corresponds to the unstable manifold of the saddle point and extends to the stable manifold tangent to one eigenvector of the stable fixed points (blue line of Fig. 1B). 
In contrast, the stable manifold of the saddle point (red line of Fig. 1B) divides the phase plane into two basins of attraction.
For large enough $\mu$, trajectories from any initial condition quickly approach close to ${\cal M}$ and then slowly evolve along ${\cal M}$ toward ${\cal A}$ or ${\cal B}$.
Timescale separation between fast and slow dynamics arises from the existence of a fast eigenvalue $\lambda_f$ and a slow eigenvalue  $\lambda_s$, where $|\lambda_f| \ll |\lambda_s|$ in the eigenspectrum of the linearized system along ${\cal M}$ (top panel of Fig. 1C). 
The slower dynamics restricted to ${\cal M}$ can be captured by a scalar field $f(x)$ where $x\propto p_B-p_A$, which is colinear to the slow eigenvector tangent to ${\cal M}(\vec{p_{0}})$ in the symmetric case.
The reduced scalar field $f(x)$ on ${\cal M}$ can be numerically computed but also approximated by a third-order polynomial by applying center manifold reduction in the neighborhood of the saddle $\vec{p}_0$ (bottom panel of Fig. 1C).
Indeed, by using appropriate linear coordinate transformations $\vec{z}=T^{-1}(\vec{p}-\vec{p_0})$, the vector field given by Eqs. 1 and rewritten as $\dot{\vec{p}}=A (\vec{p}-\vec{p}_0) + {\cal N}(\vec{p}-\vec{p}_0)$ becomes: 
\beq \label{xs1}
\frac{d\vec{z}}{dt} = J \vec{z} + \bar{\cal N}(\vec{z}) 
\eeq
where $T$ is the Jordan transformation matrix of $A$, $J=diag(\lambda_s,\lambda_f)$ and $\bar{\cal N}(\vec{z})=T^{-1}{\cal N}(T\vec{z})$. Representing ${\cal M}$ by the mapping $z_f=h(z_s)$ allows to rewrite Eq. \ref{xs1} as $\dot{z}_{s}= \lambda_s z_{s} + \bar{\cal N}_s(z_{s},h(z_{s}))$ and truncation of the taylor expansion up to third order term lead to the normal form equation:
\beq \label{xs2}
\frac{dz_{s}}{dt} = \lambda_0 + \lambda_s z_{s} - \lambda_3 z_{s}^3 + 0(z_{s}^5)
\eeq
For $\alpha=0$, $\lambda_3=p_0^2\,\mu^2(2\lambda_s-\lambda_f)-\mu/2$ is obtained by solving the invariance equation $\dot{z}_f=h'(z_s)\dot{z}_s$.
Applying the invariant manifold reduction approach in the non-symmetric case ($s_A\neq s_B$) introduces a constant symmetry-breaking term $\lambda_0 = \lambda_s (p_{A,0}-p_{B,0}) - \lambda_3 (p_{A,0}-p_{B,0})^3$ in Eq. \ref{xs2} that favors the stability of ${\cal A}$ at the expense of that of ${\cal B}$.
The scaling transformation $x=z_s/\lambda_3^{1/3}$ and the introduction of a Gaussian white noise term leads to the reduced system:
\beq
\label{1D}
\dot{x}=s + \rho\,x_i - x_i^3 + \sqrt{2D}\,\zeta_t
\eeq
where $s$ is a symmetry-breaking parameter, $\rho$ the strength of intracellular positive feeback and $D$ the variance of the noise.

\subsection*{Effective population model of inhibitory-coupled bistable cells}%

A simple population model can be built by considering $N$ cells $i$ that are described by the 1D intracellular dynamics of Eq. \ref{1D} and are coupled each other through intercellular signaling.
A common coupling term used for this general class of globally-coupled bistable systems is $\pm \gamma (x_i-x_j)$ which mediates activatory/attractive (for $-$) or inhibitory/repulsive (for $+$) coupling~\cite{Desai1978,Huber2003}. 
Furthermore, interaction delays can be incorporated as an explicit delay~\cite{Huber2003} or mediated by an intermediate mean field signal variable. From a biological viewpoint, although explicit delay is often used for models of delta-Notch signaling~\cite{Lewis2003}, a diffusive signal is assumed to be more relevant to describe paracrine signaling mediated by diffusible factors (Dif1, Fgf, CPC) or juxtacrine signaling in presence of mixing processes due to cell movement~\cite{Uriu2014}.
Accordingly, we consider the following population model:
\begin{subequations}
\begin{align}
  \frac{dx_i}{dt} & = f(x_i)  - \gamma m + \sqrt{2D}\,\zeta_{i}(t) \label{eqn1a} \\ 
  \tau_m \frac{dm}{dt} & = \frac{1}{N}\sum_{i=1,N} x_i - m   \label{eqn1c}
\end{align}
\end{subequations}
with $f(x)= s + \rho \,x - x^3$.
The aforementioned interaction $\gamma (x_i-m)$ (with $\gamma >0$) contains two terms: the first term is absorbed in the positive feedback parameter $\rho$ and thus contributes to bistability~\cite{Matsuda2015} while $-\gamma m$ mediates a global negative feedback of timescale $\tau_m$ related with the time-consuming processes of synthesis, degradation, regulation or diffusion of the signaling components.
For the following, it is convenient to define the parameterized potential,
\beq
U(x,m)=\int_{\mathbb{R}} \left[ f(x) - \gamma m \right]\,dx
\label{U}
\eeq
which exhibits two wells $x_j(m)$ ($j=\cal A$ or $\cal B$) and a saddle $x_{\cal S}(m)$ for $\gamma m \in [r_1,r_2]$ with basins $\Omega_j$, curvatures $U''_{\cal A,\cal B,\cal S}(m)$ and barrier heights $\Delta_{\cal A,\cal B}(m)$.
The Kramers escape rate from the well $j=\cal A$ or $\cal B$ is given by:
\beq
r_{j}(m) = \frac{\sqrt{U''_{j}(m)|U''_{\cal S}(m)|}}{2\pi}\,\exp{\left(-\frac{\Delta_{j}(m)}{D}\right)}.
\label{kram}
\eeq

\section*{Results}%

\subsection*{Steady-state proportion between two cell-type clusters}

To investigate the dynamic and steady-state properties of the population model (Eqs. 5), it is convenient to consider the continuum limit $N \ra +\infty$ for which the model can be reformulated in terms of probability distribution function $P(x,t)$:
\begin{subequations}
\begin{align}
  \frac{\partial}{\partial t}P & =\frac{\partial}{\partial x}\left(\frac{d}{dx}U(x,m)\,P\right)+D\frac{\partial^2}{\partial x^2}P \label{fp1} \\
  \tau_m\,\frac{dm}{dt} & = \int_{\mathbb{R}} x\,P(x,t)\,dx-m 
\end{align}
\end{subequations}
The stationary solutions of Eqs. 8 satisfy the self-consistent equation~\cite{Desai1978}:
\beq
m=\int_{\mathbb{R}} x\,{\cal P}(x,m)\,dx \equiv q(m) 
\label{hm}
\eeq
where ${\cal P}(x,m) = {\cal N}(m)^{-1}\,\exp{\left(-U(x,m)/D\right)}$ (${\cal N}$ is the normalization prefactor).
For $\gamma>0$, Eq. \ref{hm} has a unique stationary solution $\bar{m}$ that is associated with a bimodal steady-state distribution $\bar{{\cal P}}(x)={\cal P}(x,\bar{m})$ for $\rho$ large enough and $s$ small enough.
Several other steady-state quantities can be defined the same way: potential $\bar{U}(x)$, fixed points $\bar{x}_{j}$, attraction basins $\bar{\Omega}_{j}$, Kramers escape rates $\bar{r}_{j}$, cluster sizes $\bar{{\cal P}}_{j}=\int_{\bar{\Omega}_{j}} \bar{{\cal P}}(x)\,dx$ and cluster proportion $\bar{R} = \bar{{\cal P}}_{\cal A}-\bar{{\cal P}}_{\cal B}$.
This two-cluster steady state  $\bar{m}$ is stable for any $\gamma$, $D$ and $\tau_m$ values, which can be demonstrated in some limits by computing the lowest stability exponent.
In the limit $\tau_m \rightarrow +\infty$, the probability distribution follows adiabatically the slow population dynamics and the stability of the two-cluster steady state is determined by the eigenvalue $q'(\bar{m})-1$ (Eq. \ref{hm}) that is always negative for inhibitory coupling as $q'(\bar{m})<0$ for $\gamma>0$.

For noise small enough so that the escape time from one well is large compared with the intrawell relaxation time, the slow time evolution of the mean field $m$ and proportion $R={\cal P}_{\cal A}-{\cal P}_{\cal B}$ depends on Kramers escape rate $r_j(m)$ (Eq. 7) and well position $x_j(m)$ as, 
\begin{subequations}
\begin{align}
  \frac{dR}{dt}  &= r_{\cal B}(m)-r_{\cal A}(m)-(r_{\cal A}(m)+r_{\cal B}(m))R \label{drdt} \\
  \tau_m\frac{dm}{dt} &= (x_{\cal A}(m)(1+R)+x_{\cal B}(m)(1-R))/2-m
\end{align}
\end{subequations}
The steady state satisfies $\bar{R} =  \frac{2\bar{m}-\bar{x}_{\cal A}-\bar{x}_{\cal B}}{\bar{x}_{\cal A}-\bar{x}_{\cal B}} \equiv g(\bar{m})$ and the eigenvalues of the linearized system around $\{\bar{m},\bar{R}\}$ are found to be always negative for inhibitory coupling as $\bar{r}'_{\cal A}(m)>0$ and $\bar{r}'_{\cal B}(m)<0$ for $\gamma >0$.

The cell-type proportioning process requires that a broad range of initial conditions converges to the two-cluster steady state, but also rapidly enough with respect to the developmental time $\tau$ ($\gg \{1,\tau_m\}$) and stable enough with respect to noise-induced interwell hopping.
In fact, we show the existence of three distinct regimes that may obstruct efficient proportioning (Fig. 2): (i) frustrated relaxation that compromises precise proportions in finite time; (ii) steady-state hopping that hinders robust cell-type acquisition against noise; (iii) collective oscillations that compromise stable proportions over time. 
These regimes occur for specific ranges of parameters (Fig. 2A,B) and are illustrated for the extreme case of an initial condition where all cells start in a state strongly shifted to high $x$ levels and biased toward ${\cal A}$ fate (Fig. 2A-F).

\subsection*{Trade-off between precise proportion and robust commitment}

For $D$ low enough and $\tau_m$ not too large, the relaxation rate toward a steady state $\{\bar{m},\bar{R}\}$ is governed at long time by the slowest relaxation modes determined by the Kramers transition rates.
At this slow timescale assumed to be larger than $\tau_m$, $m$ adiabatically follows the change of proportion $R(t)=\int_{\Omega_{\cal A}} P(x,t)dx-\int_{\Omega_{\cal B}} P(x,t)dx$ which is driven by the Kramers transition rates according to Eq. \ref{drdt} with $m=g^{-1}(R)$.
The rate of the monotonous perturbation decay $\delta R(t)=|R(t)-\bar{R}|$ is mostly proportional to the largest Kramers escape rate that typically decreases by several orders of magnitude as $\delta R$ diminishes. 
As a result, the relaxation can become critically slow (i.e., frustrated) at the timescale $\tau$ when the system has reached a critical distance from the steady state.
Given that the mapping ${\cal H}(\delta R(t_0))=\delta R(t_0+\tau)$ has a decreasing derivative function (constant derivative for exponential decay), one can define a maximum frustrated proportion $\delta R_\epsilon(\vec{p})$ satisfying:
\beq
{\cal H}(\delta R_\epsilon,\vec{p})=\delta R_\epsilon\,(1-\epsilon)
\label{frust}
\eeq
where $\epsilon \ll 1$ and $\vec{p}=\{D,\gamma,s,\rho\}$. 
For $s \ne 0$, such frustration level is different for positive and negative perturbation and needs to be normalized as $\delta R_\epsilon/(1+|\bar{R}|)$.
If ${\cal H}(\delta R_{\epsilon}(m),\vec{p})-(1-\epsilon)$ is positive or negative for $m \in [r_1/\gamma,r_2/\gamma]$, then $\delta R_{\epsilon}(\vec{p})=0$ or $\delta R_{\epsilon}(\vec{p})=1$, respectively.
Frustration level $\delta R_{\epsilon}$ is the highest for $D$ and $\gamma$ small and $s$ large (Fig. 2A,B,C).
Increasing $\gamma$ diminishes frustration by narrowing the range of $m$ values associated with bistability ($m \in [r_{1}/\gamma,r_{2}/\gamma]$) and, therefore, by restricting the possible range of $\delta {R}_{\epsilon}$.
Higher noise (or lower barrier height) also reduces frustration by increasing the perturbation decay rate as $r \propto \exp{(-\Delta/D)}$ (Fig. 2D).

Although noise-induced interwell hopping is beneficial to reduce frustration during relaxation, it is biologically unlikely when equilibrium is reached as stochastic transdifferentiation is not observed and harmful in fully developed organisms.
We thus define a biologically-irrelevant hopping regime (Fig. 2A,B) for which non-negligible noise-induced interwell hopping occur during time interval $\tau$ (Fig. 2E), which occurs for a noise larger than $D_H$ given by,
\beq
(\bar{R}+1)\,\bar{r}_{\cal A}(D_H)  \, \tau =1 .
\label{NS}
\eeq
Note that the hopping regime necessarily coincides with an absence of frustration.
A trade-off between competing requirements for noise-driven relaxation and noise-robust clusters restricts the occurrence of efficient proportioning to a narrow and specific range of noise level.

\subsection*{Collective oscillations through global negative feedback}

Although increasing the strength of inhibitory coupling reduces the frustration level $\delta R_{\epsilon}$, it can also give rise to collective oscillations in presence of large enough coupling delays. 
Some hints regarding the stability of these collective oscillations can be gained by investigating the model without noise ($D=0$) and with identical elements ($x_i \rightarrow x$) without noise ($D=0$):
\begin{subequations}
\begin{align}
\frac{dx}{dt} &= s - \gamma m+\rho\,x-x^3 \\
\tau_m\frac{dm}{dt} &= x - m .
\label{sm5}
\end{align}
\end{subequations}
This class of two-dimensional dynamical system has been studied in details by Boissonade and de Kepper (1980).
The nullcline $\tilde{x}(m)$ obtained by solving $f(x)-\gamma m=0$ has a Z shape for $m \in [r_1/\gamma,r_2/\gamma]$, such that stable hysteretic oscillations occur for large enough coupling strength and delay for which all the fixed points of the deterministic system are unstable. 
For $s=0$, there is a single fixed point ($x=0$ and $m=0$) for $\gamma>\rho$, and such fixed point is unstable for $\tau_m>1/\rho$. 
Stable oscillations occurs in presence of a single unstable fixed point when inhibitory coupling is strong enough ($\gamma>\rho$) and slow enough ($\tau_m>1/\rho$), but also in presence of three unstable fixed point for $\gamma<\rho$.
In presence of moderate level of noise, numerical simulations show the existence of a stable one-cluster oscillatory state that coexists with the two-cluster steady state and is destabilized toward this state above a critical noise level $D_{O}(\vec{p})$ ($<D_H$) (Fig. 2A,E).
Note that such oscillatory behavior quite differs from the case of explicit delays for which the two-cluster steady state can be destabilized toward multiple oscillatory states~\cite{Huber2003}.

\subsection*{A dual-time positive feedback improves proportion regulation}

Population of inhibitory-coupled bistable cells exhibits a stable two-cluster state that nevertheless tends to be either too unstable at single-cell level (hopping regime) or weakly attracting at population level (collective frustration or oscillations).
To resolve this antagonism, we propose a solution based on the existence of multiple regulatory timescales, which allows to control separately the relaxation and the steady-state properties.
Indeed cellular differentiation relies not only on transcriptional positive feedback loops but also on slower positive feedback loops for instance mediated by epigenetic mechanisms that stabilize gene expression pattern~\cite{Dodd2007,Furusawa2013,Sasai2013}.
To describe such slower reinforcement mechanism, the model of the intracellular dynamics (Eq. 5a) is supplemented with a slow variable $y_i$ as follows:
\begin{subequations}
\begin{align}
\frac{dx_i}{dt} & = f(x_i) +  \sqrt{\frac{\beta}{\tau_y}}\,y_i - \gamma m + \sqrt{2D_x}\,\zeta_{x,i}(t) \label{eqn10a} \\ 
\tau_y \frac{dy_i}{dt}   & = \sqrt{\beta\tau_y} \,x_i -y_i + \sqrt{2D_y\,\tau_y}\,\zeta_{y,i}(t) \label{eqn10b}  
\end{align}
\end{subequations}
where $\tau_y$ and $\beta>0$ are respectively the timescale and the strength of the so-called epigenetic feedback.
The choice of linear coupling terms between variables $x$ and $y$ and the further assumption that $D_x = D_y \equiv D$ are not critical but very convenient as a two-dimensional potential $U(x,y,m)=U(x,m)+\sqrt{\beta/\tau_y}\,\,x\,y-\frac{1}{2\,\tau_y}y^2$ can be defined and stationary solutions $\bar{{\cal P}}(x,y)$ and $\bar{m}$ can be obtained by replacing $U(x,m)$ with $U(x,y,m)$ in Eq. \ref{fp1}.
Using a timescale separation argument, an epigenetic feedback that is slow enough ($1 \ll \tau_y < \tau$) and inactivated at early time ($y(0) \sim 0$) allows to reduce the occurrence of frustrated relaxation, steady-state hopping and collective oscillations (Fig. 3).
On the one hand, frustration regime is mostly determined at early time by transition rates associated with the fast system and potential $U(x,0,m)$ whose saddle barrier can be lowered by decreasing $s$.
On the other hand, the hopping regime is determined by the steady-state potential $\bar{U}(x,y)$ whose saddle barrier can be heightened by increasing $\beta$. 
Furthermore, oscillations are precluded in the fast system due to the low ratio $\Delta(s)/D$ (for $s$ small) as well as in the slow system if one assume a slower timescale for the epigenetic positive feedback than for the global negative feedback (for $\tau_y>\tau_m$).
As a result, dual fast-slow positive feedback can significantly extend the domain of efficient proportioning by controlling separately and thus reducing simultaneously the respective domains of frustration, hopping and oscillations (Fig. 3A). 

The efficiency of this proportioning mechanism therefore relies on the existence of two relaxation phases well-separated in time (Fig. 3B,C). 
At short time scale, fast intracellular feedback, noise level and intercellular coupling strength contribute to a full relaxation of the fast modes toward the steady state associated with low $y$ values and low barrier height, while further activation of the slow epigenetic feedback stabilizes both cell fates and cluster size proportions. 
Although introducing a slow feedback is critical for stabilizing cell-fate decisions, keeping a fast positive feedback mechanism is still required for rapid symmetry-breaking and cell-fate diversification, otherwise the cell-type proportioning would occur at the time scale of the slow positive feedback.
This biphasic relaxation mechanism can operate to set any final proportion $R(\tau)$ value between $0$ (equal proportion) and $1$ (all-or-none proportion) by tuning $s$ (Fig. 3D).
A minor caveat is that $\bar{R}$ depends on $D$ (in proportion to $s$) while an initial epigenetic bias may lead to significant proportion errors associated with frustration $\delta R(\tau)$.

Overall, this proportioning mechanism requires a specific hierarchy of the system time scales where $1/\bar{r}|_{\bar{U}(x,y)} > \tau \sim \tau_y > 1/\bar{r}|_{\bar{U}(x,0)} >\tau_x \sim \tau_m $, which is consistent with the typical biological time scales ranging from less than an hour for intercellular signaling~\cite{Herrgen2010}, hours for protein expression changes through synthesis and degradation processes, few hours to days for epigenetic regulatory events~\cite{Sasai2013,Durso2014} and days for the developmental time scale.

\section*{Discussion}%

Using a generic modeling framework, we have identified a minimal set of mechanisms required to perform an efficient proportion control of distinct differentiated cell types emerging during multicellular development.
In population of interacting bistable cells, noise and inhibitory coupling need to be finely tuned to reach a collective state characterized with precise proportion between robust cell types. 
Fine-tuning is due to the antagonistic requirements between the respective tasks of diversification, proportioning and stabilization of cell-state attractors. 
Stochasticity is important to create small differences between identical cells~\cite{Losick2008,Meyer2014} and to escape from metastable cell states, but is detrimental for robust cell-type specification.
For its part, lateral inhibition contributes to amplify small cellular differences~\cite{Losick2008,Matsuda2015} and is critical for rapid relaxation within a narrow proportion range, whereas it promotes oscillatory behaviors or becomes counterproductive after cell sorting occurs.
A solution to these antagonisms consists in the existence of dual positive feedback loops operating at distinct timescales, which coordinately orchestrate in time the emergence of diverse cell types, the precise regulation of their proportion and their robust and quasi-irreversible fate commitment.

The importance of a multiphasic differentiation process is supported by the common observation of a lineage priming phase before the irreversible lineage-restricted fate commitment.
This developmental stage is typically characterized with dynamic expression of differentiation factors during which pre-commitment decisions are reversible, such as in Amoeba~\cite{Tasaka1983,Chattwood2011} or mammal embryos~\cite{MartinezArias2011,MartinezArias2013}. 
In this process, the cooperation between transcriptional noise and feedback is critical to provide flexible fate switching abilities and quickly reach steady-state proportions.
The further transition from reversible lineage priming to irreversible fate determination happens to involve the activation of a delayed positive feedback mechanism, which is most likely mediated by epigenetic regulatory processes~\cite{Hemberger2009} but also by the activation of autocrine signaling pathways~\cite{Anjard2005} or by the spatial segregation of cell subpopulations~\cite{Xenopoulos2015}. Importantly, our model suggests the possibility to autonomously schedule this transition between reversible and irreversible commitment, as late epigenetic activation occurs when single-cell dynamics has been stabilized to some extent, thereby reflecting that some steady-state proportion has been reached. 
Although this biphasic differentiation dynamics can establish precise, symmetric or asymmetric, proportions for a wide range of initial conditions and of noise levels, it remains sensitive to two classes of perturbations.
On the one hand, parametric perturbations of signaling or intracellular dynamics (parameter $s$ in the model studied here) can affect the proportions, as it has been shown when manipulating extracellular levels of Dif1 in Dictyostelium~\cite{Parkinson2011} or of Fgf4 in mammalian embryos~\cite{Yamanaka2010,Krawchuk2013}.
On the other hand, transient perturbations of proportion after cell removal or death during the late stabilization phase can lead to frustration, which accounts for the error tolerance zone observed in Dictyostelium~\cite{Rafols2001} or for the use of a pool of undifferentiated stem cells to restore proportions in plants or animals~\cite{Birnbaum2008}.

Other sophisticated regulatory processes influences cell-fate decisions and, indirectly, the proportioning process.
Notably, stem or progenitor cells are subjected to an highly dynamic control of differentiation factors, through ultradian oscillations~\cite{Imayoshi2013}, cell-cycle progression~\cite{Pauklin2013}, or asymmetric division~\cite{Zhong2008}.
All these processes are prone to contribute to dynamic cellular heterogeneity in term of cell-fate propensity, which further promotes both rapid and divergent fate decisions and thus minimizes the risk of too slow and frustrated differentiation without requiring other sources of stochasticity~\cite{Furusawa2001,Pfeuty2015}.
Although the intracellular mechanisms that switch complex oscillatory dynamics toward diverse steady states can be very complicated, low-dimensional models can nevertheless be used to address these issues~\cite{Ullner2007,Pfeuty2014}.

The developmental regulation of cell-type proportion mechanisms also uses mechanisms involved, in the first place, in tissue growth and patterning. 
The fact that fate-restricted progenitor cells keep the ability to proliferate suggests that proportioning can also be controlled through the relative proliferation rates rather than interconversion rates between distinct types of progenitors~\cite{Kicheva2014}.
Finally, spatial regulation of cell-type specification adds a whole dimension to the issue of proportioning. 
Spatial arrangement of cell types into domains often occurs after the proportioning process through cell motility and sorting process~\cite{Nicol1999,Xenopoulos2015}, but it may also take place simultaneously through a spatial control of fate decision depending on positional informations relative to compartment's boundaries and signaling centers~\cite{Wolpert1969,Ralston2008}. 
Yet, the presence of morphogen gradient does not preclude the need for lateral signaling to regulate domain size and sharpen domain boundary against various sources of noises~\cite{Xiong2013}.

\section*{Conclusion}%

This study highlights how the developmental establishment of diverse cell types requires a well-orchestrated interplay between intracellular, intercellular and stochastic mechanisms.
The challenge lies in reconciling the competing demands of flexible decisions during early cell-type diversification and proportioning and of robust lineage-restricted fate commitment for specialization purposes.
The existence of multiple regulatory timescales appear critical to dissociate the positive feedback mechanisms that cooperate with noise and lateral inhibition to promote cellular heterogeneity and tune proportions, to those required to lock fate decisions regardless intercellular signals and noise levels.

\section*{Acknowledgments}%

We thank Nen Saito for stimulating discussions. 
BP acknowledges support of CANON foundation in Europe fellowship. 
KK acknowledges support of MEXT Japan.

\bibliographystyle{biophysj}
\bibliography{fateratio}

\clearpage

\begin{figure}
  \centering
  \includegraphics[width=8.9cm]{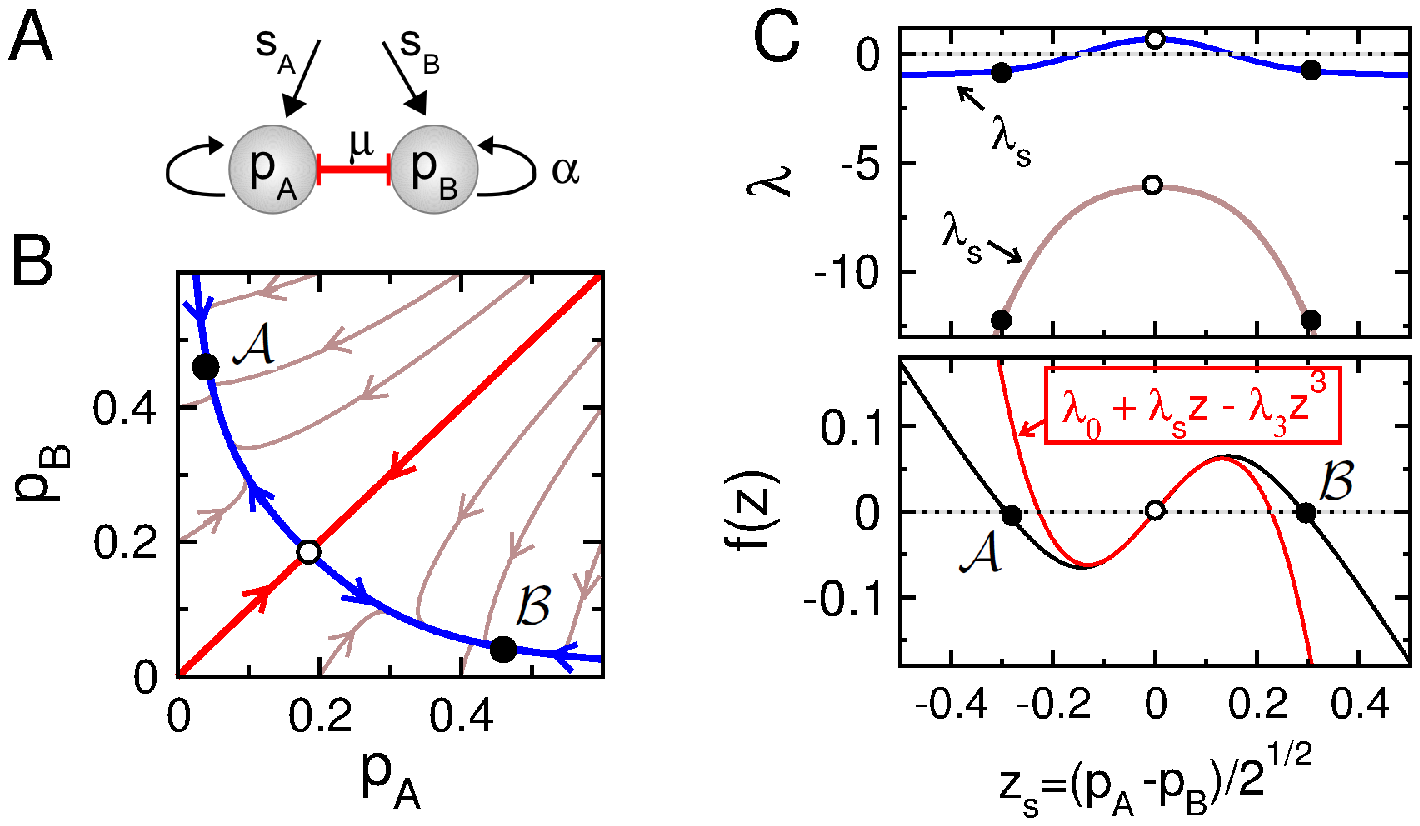}
  \caption{{\bf One-dimensional model of bistable cell dynamics.}
(A) Schematic representation of a typical positive-feedback circuit involved in regulating cell differentiation. black/red lines: positive/negative interaction.
(B) Phase portrait depicting the fixed points (circles), the separatrix (red), the invariant slow manifold ${\cal M}$ (blue) and example of trajectories (brown). Parameters: $\mu=50$, $s_{A,B}=0.5$ and $\alpha=0$.
(C) The invariant manifold ${\cal M}(z_s)$ is characterized by fast and slow eigenvalues $\lambda_f$ and $\lambda_s$ of the Jacobian (top panel) and by a flow velocity (black line of bottom panel) approximated by a third-order polynomial function given by Eq. 3 (red line of bottom panel).}
  \label{fig1}
\end{figure}

\clearpage

\begin{figure}
  \centering
  \includegraphics[width=8.9cm]{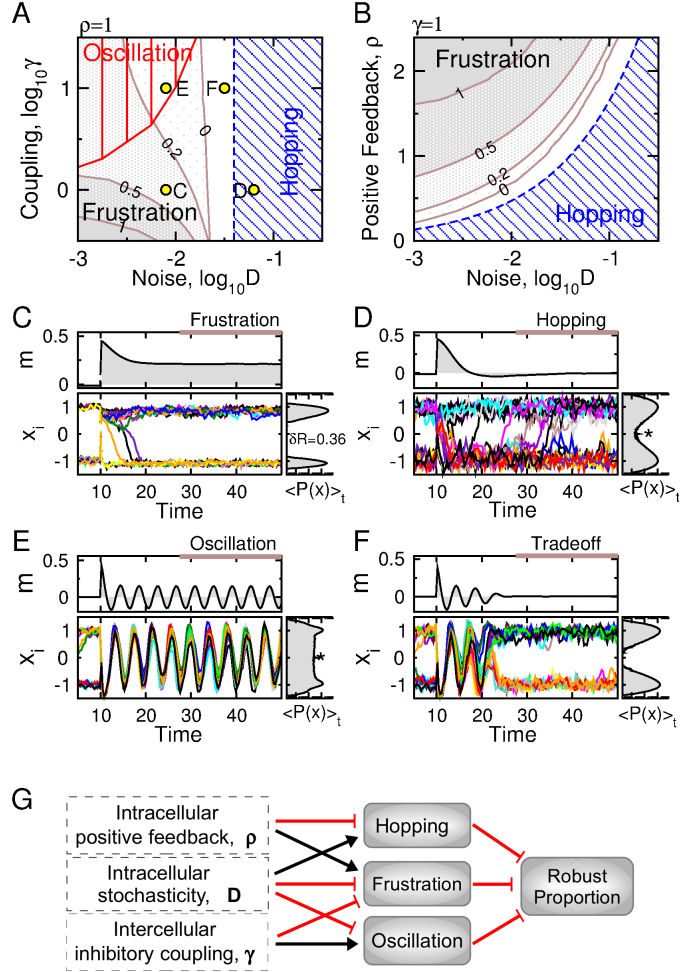}
  \caption{\noindent {\bf Frustrated relaxation, steady-state hopping and collective oscillations.}
    Parameters: $s=0$, $\tau_m=4$, $\rho=1$ and $\tau=1000$ or otherwise indicated.
    (A-B) Parameter domains associated with stable oscillations (red vertical hatch), frustrated relaxation (dotted domains where brown boundaries are isolines $\delta R_{0.9}$ given by Eq. 11) and steady-state hopping (blue diagonal hatch delimited by blue line given by Eq. 12). Indexed circles correspond to examples shown in panels C to F.
    (C-F) Simulated individual and mean-field trajectories in response to a perturbation of the two-cluster steady state for each regime (circles in panel A) and time-averaged probability distribution $\langle P(x,t) \rangle_{[\tau-T,\tau]}$.
    (G) Schematic representation of the existence of competing mechanisms and regimes, which complicates the task of robust proportioning.}
  \label{fig2}
\end{figure}

\clearpage

\begin{figure}
  \centering
  \includegraphics[width=8.9cm]{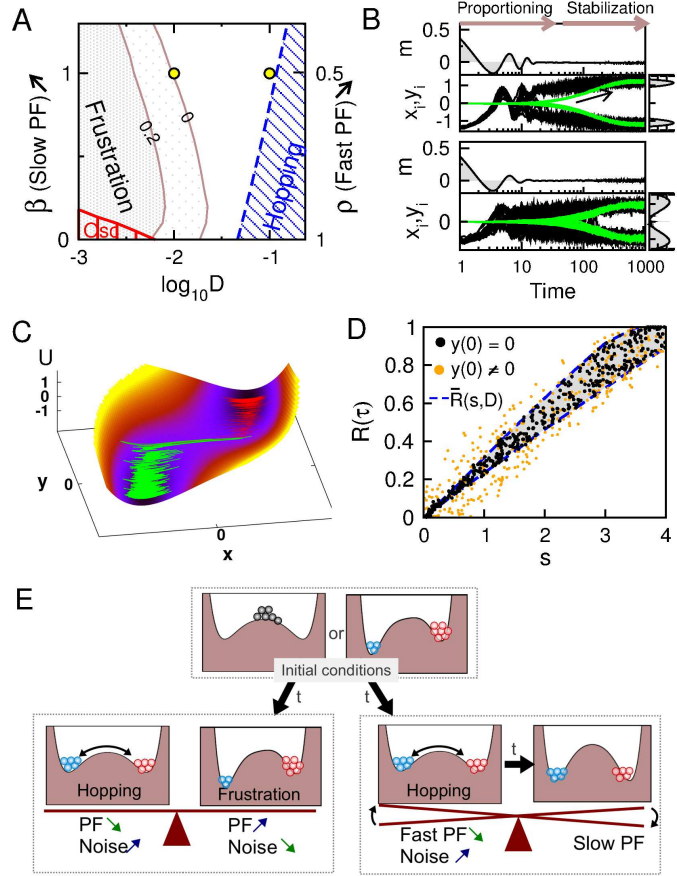}
  \caption{{\bf Efficient proportion regulation: dual positive feedback and biphasic relaxation.}
    Parameters: $s=0$, $\tau_m=4$, $\gamma=5$, $\tau_y=100$, $\tau=1000$, $\rho=0.5$, $\beta=1$ or otherwise indicated.
    (A) Frustration, oscillatory and hopping domains (see Fig. 2 for legend) as function of $D$ and $\beta$ (with $\rho=1-\beta/2$).
    (B) Relaxation dynamics of $m$, $x_i$ and $y_i$ represented in time (up: $D=10^{-2}$; bottom: $D=10^{-1}$).
    (C) State-space trajectories of individuals schematically drawn on the steady-state potential $\bar{U}(x,y)$.
    (D) Distribution of final proportion $R(\tau)$ as a function of $s$ for uniform distribution of $\log_{10}D \in [-2,-1]$, $x_i(0) \in [-1,1]$ $m(0)\in [-1,1]$, for $y_i(0)=0$ or $y_i(0)\in [-\sqrt{\beta\tau_y},\sqrt{\beta\tau_y}]$. 
    Dashed lines: $\bar{R}$ computed for $D=10^{-2}$ and $D=10^{-1}$, respectively.
    (E) Schematic representation of relaxation dynamics on a bistable potential, $U(x)$, from out-of-equilibrium initial conditions: 
    Slow positive feedback avoids the regimes of hopping and frustration by monitoring in time the balance between intracellular positive feedback and noise.}
  \label{fig3}
\end{figure}

\end{document}